\documentclass[journal]{IEEEtran}
\usepackage{url}
\usepackage{caption}
\usepackage[noadjust]{cite}
\usepackage{graphicx}
\usepackage{multirow}
\usepackage{amsmath}
\usepackage{subcaption}
\usepackage{comment}
\usepackage{makecell}
\usepackage{ragged2e}
\usepackage{rotating}
\usepackage{lscape}
\usepackage[table]{xcolor} 
\usepackage{nomencl}
\usepackage{stfloats}
\makenomenclature
\usepackage{wrapfig}
\usepackage{makecell}
\usepackage{balance}
\usepackage[utf8]{inputenc}
\usepackage{ragged2e}
\usepackage{booktabs}

\usepackage{pifont}
\usepackage{tabulary}
\usepackage{comment}
\usepackage{lipsum}
\usepackage{array, tabularx, multirow, booktabs, diagbox}
\usepackage{adjustbox}

\usepackage{amsthm}

\theoremstyle{remark}

\begin{document}
\title{Large Language Models-Empowered Wireless Networks: Fundamentals, Architecture, and Challenges}
\author{Latif~U.~Khan,~\IEEEmembership{Member,~IEEE},~Maher~Guizani,~Sami~Muhaidat,~\IEEEmembership{Senior~Member,~IEEE},~Choong~Seon~Hong,~\IEEEmembership{Fellow,~IEEE}
        
    
\IEEEcompsocitemizethanks{
\IEEEcompsocthanksitem L.~U.~Khan~is with the Department of Computer Science \& IT, Abu Dhabi University, United Arab Emirates. 
\IEEEcompsocthanksitem M.~Guizani is with the Computer Science and Engineering Department, University of Texas at Arlington, USA.
\IEEEcompsocthanksitem S. Muhidat is with the $6$G research center, Khalifa University, United Arab Emirates.
\IEEEcompsocthanksitem C.~S.~Hong is with the Department of Computer Science \& Engineering, Kyung Hee University, South Korea. 

\IEEEcompsocthanksitem Corresponding author: Latif U. Khan [latif.u.khan2@gmail.com]

}}

\markboth{IEEE Internet of Things Magazine}{}%

\maketitle





\begin{abstract}  
The rapid advancement of wireless networks has resulted in numerous challenges stemming from their extensive demands for quality of service towards innovative quality of experience metrics (e.g., user-defined metrics in terms of sense of physical experience for haptics applications). In the meantime, large language models (LLMs) emerged as promising solutions for many difficult and complex applications/tasks. These lead to a notion of the integration of LLMs and wireless networks. However, this integration is challenging and needs careful attention in design. Therefore, in this article, we present a notion of rational wireless networks powered by \emph{telecom LLMs}, namely, \emph{LLM-native wireless systems}. We provide fundamentals, vision, and a case study of distributed implementation of LLM-native wireless systems. In the case study, we propose a solution based on double deep Q-learning (DDQN) that outperforms existing DDQN solutions. Finally, we provide open challenges.
\end{abstract}


\section{Introduction}
\textcolor{black}{In the upcoming years, we will witness novel applications (e.g., metaverse and holographic applications, i.e., healthcare, intelligent transportation systems, brain-computer interaction, and industry 5.0, among others) that will be very difficult to enable using the existing wireless systems \cite{saad2019vision}. Therefore, we will need a transition from traditional architecture to novel architecture/s. This transition is due to the novel's diverse requirements (e.g., metaverse with human-like decisions and semantic communication with logical reasoning abilities) for novel use cases. Therefore, for effective enabling of such kinds of applications, there is a need for novel system design that follows new design trends. Mainly, these design trends are self-organizing and proactive analytics. \emph{Self-organizing} trend enables wireless networks to operate autonomously, i.e., as to optimize their parameters, configure, and adjust autonomously. The design trend of \emph{Self-organizing} is necessary due to the reason of expecting massive devices and entities with a wide variety of requirements in the foreseeable future. On the other hand, \emph{proactive analytics} is necessary to analyze the system prior to user requests so as to serve them instantly when they request.     }\par
\textcolor{black}{On the other hand, machine learning (ML) has proven to be one of the key enablers of emerging wireless networks/Internet of Things (IoT) \cite{khan2023joint}. Various applications of ML in IoT are smart homes energy management, predictive maintenance in Industry $4.0$, wireless radio resource management, intelligent sensing for metaverse applications, and lane change assistance in autonomous driving cars, among others. The notion of applying ML in IoT is due to the widespread data generated by many IoT applications devices and the difficulty in using traditional mathematical modeling schemes for various complex scenarios (e.g., solving non-convex, NP-Hard complex wireless problems). For instance, mobility modeling of mobile devices in a wireless system is challenging due to its highly dynamic nature. To do so, one can use mathematical modeling based on a random way-point model, random walk model, and Gauss-Markov mobility model, among others. Although these models perform well but might not be able to exactly model the highly dynamic mobile nodes. To address this, one can use deep learning-based mobility modeling \cite{luca2021survey}. Similarly, there are many scenarios in IoT systems that can be modeled using machine learning. However, training an effective machine learning model requires a significant amount of training data that is not always easily available. Therefore, there is a need for data generation to enable effective and efficient training of machine learning models. Meanwhile, there will be novel applications that will need analysis prior to deployment. For analysis of such kinds of applications, we will need data as well as virtual scenarios. Furthermore, there are scenarios (e.g., healthcare applications) where the training data is limited. Other than that, there might be scenarios (e.g., semantic communications), where we need logical reasoning that is very difficult using traditional ML. To address these challenges, there is a need for novel ML schemes. Such schemes could be based on large language models (LLMs). }\par      
\textcolor{black}{Some works considered LLMs and wireless \cite{bariah2023large, maatouk2024large, zou2024telecomgpt, karapantelakis2024using}. The work in \cite{bariah2023large} presented the use of LLMs for telecom applications. They discussed key motivations followed by the use cases. Another work \cite{maatouk2024large} discussed the fundamentals of LLMs for telecom with a case study of 3GPP document comprehension. The work in \cite{zou2024telecomgpt} focused mainly on dataset collection and then tested it. The other work \cite{karapantelakis2024using} analyzed existing LLMs for question/answering telecom documents. Different from the works in \cite{bariah2023large, maatouk2024large, zou2024telecomgpt, karapantelakis2024using}, we propose a novel telecom LLM agents-based architecture for effectively enabling emerging wireless applications. Furthermore, we provide a novel joint distributed LLM learning and task offloading framework and propose a solution based on double deep Q-learning (DDQN) along with convex optimization for time interval allocation and relative local accuracy. Our contributions are summarized as follows: }\par
\begin{itemize}
    \item We discuss the fundamentals of LLMs for wireless networks. Specifically, we discuss the limitations of existing wireless networks enabled by traditional ML. Then, we provide the notion of LLMs in enabling wireless networks.
    \item We present a vision of telecom LLM agents for effectively enabling wireless applications. Furthermore, we identify three main aspects of telecom LLM agent-based design and discuss their significance in detail.
    \item We present a case study of joint distributed LLM agents learning and task offloading. For a solution, we consider DDQN and modify it for faster convergence.
    \item Finally, we present open challenges and conclude the article.
\end{itemize}

\begin{figure*}[!t]
	\centering
	\captionsetup{justification=centering}
	\includegraphics[width=16cm, height=16cm]{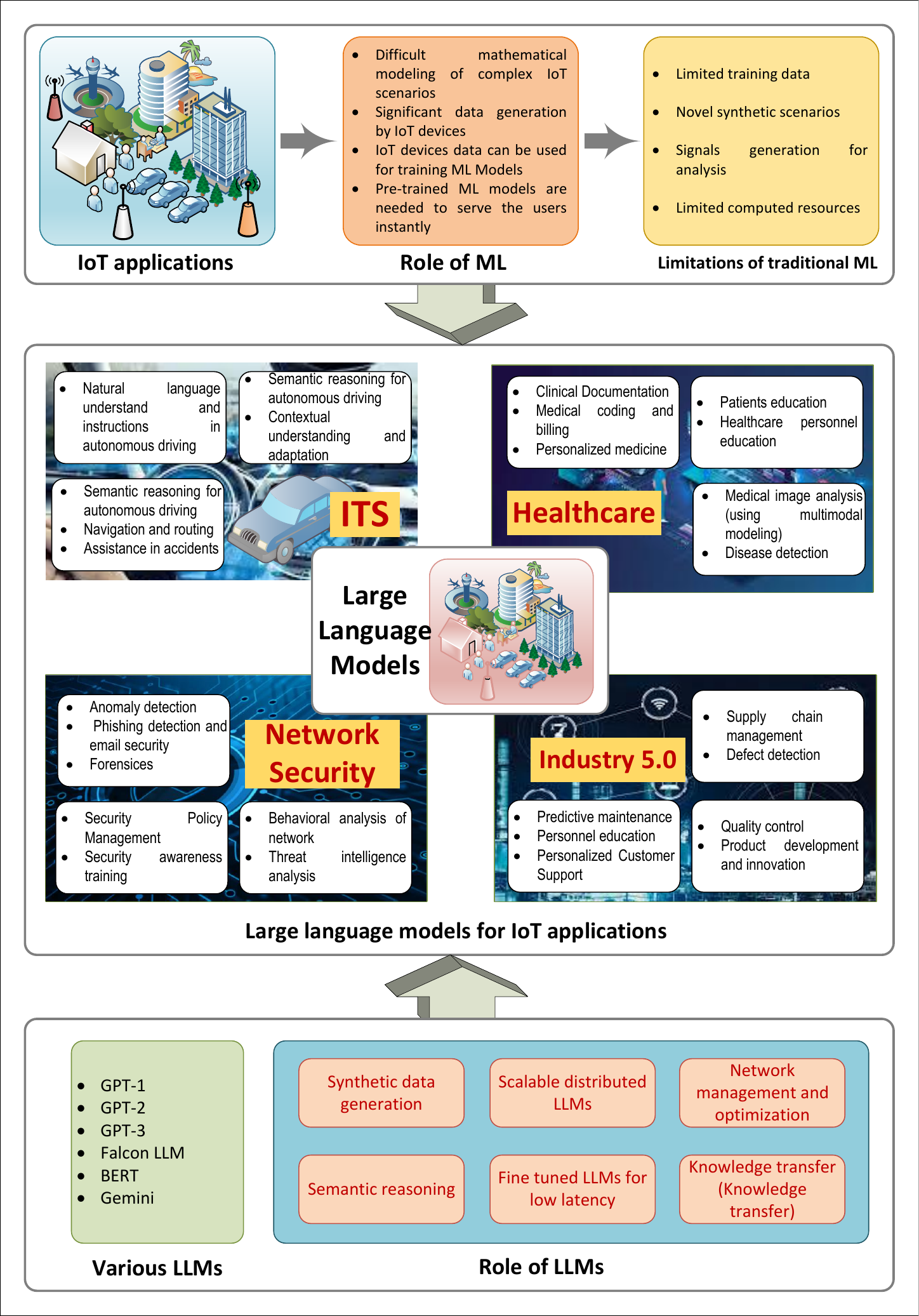}
	\caption{An overview of large language models for IoT.}
	\label{fig:intro}
\end{figure*}

\section{Large Language Models for Rational Wireless Networks: Motivation, Foundations, and Vision}
\subsection{Limitations of Traditional Wireless Systems}
\textcolor{black}{Mathematical optimization has been widely used for modeling of a wide variety of applications/functions in wireless networks (e.g., resource allocation and association). However, Wireless systems encompass a diverse array of applications and functions, making the implementation of mathematical optimization strategies challenging \cite{saad2019vision, khan2023network}. For example, consider the scenarios of brain-computer interaction, resource optimization, and task-offloading for edge networks, among others. In brain-computer interaction, a flawless interface between end devices and the human brain is essential. For instance, in healthcare applications, prosthetic limbs can be controlled through brain implants by humans. Meanwhile, researchers are exploring new opportunities to enable many novel applications, ranging from brain-controlled autonomous cars to brain-controlled entertainment in the cinema. Based on brain-computer interaction, there is a need for the use of various implants and devices for users to communicate with their environment. End-users can use gestures to communicate with other devices and people. Such interaction-based empathic and haptic communications will lead to many design challenges that are very difficult to be modeled by traditional wireless systems, as shown in Fig.~\ref{fig:intro}.} \textcolor{black}{Similar to brain-computer interaction, resource optimization for emerging applications and task offloading in the edge metaverse need careful attention. It is very difficult to simultaneously optimize all variables (e.g., wireless resource allocation, computing resource optimization, and association) and minimize the cost function (e.g., QoE) for wireless networks. }\par

\subsection{Limitations of Traditional ML-Enabled Wireless Systems}
\textcolor{black}{While ML can facilitate a diverse range of wireless applications, certain hurdles must be confronted.} \par

\begin{itemize}
    \item \textit{Insufficient Training Data:} \textcolor{black}{There are a wide variety of applications (e.g., healthcare, remote sensing, and autonomous systems) where the training data might be limited and thus making it hard for ML models to train. In digital healthcare, it is very challenging for collecting data due to the less patients for a rare disease, privacy concerns where the patients don't want to share their data with the authorities, and difficulties in collecting data from the wearable. On other other hand, in smart agriculture, the data collection is mainly challenging due to insufficient sensory infrastructure. Similarly, collecting data in live animals is also challenging due to animal behavior and expensive sensory infrastructure. Meanwhile, the limited availability of data will result in many issues. These issues are limited representation, generalization issues, overfitting, class imbalance, and feature extraction issues. Based on the aforementioned discussion, it is necessary to generate more data for obtaining effective ML models.}     
    
   \item \textit{Synthetic Scenarios for Novel Applications:} \textcolor{black}{To deploy novel wireless applications, there can be many ways. One way is to implement a wireless system without analyzing the virtual model of the physical system. This approach is costly and might not prove very effective due to the fact that the deployed system might not be able to fulfill the end-user's requirements. Another way is to perform simulations that mimic the real-world scenario. For this kind of analytics, there is a need for synthetic data that closely resembles real-world data for training machine learning models. One can use large language models for data generation for such kinds of scenarios. }  
    \item \textit{QoE Constraints:} \textcolor{black}{The ability of conventional machine learning (ML) models to give chatbots a smooth and interesting conversational experience is constrained. These models frequently rely on manually created characteristics and established rules, which limits their capacity to comprehend intricate linguistic structures, sophisticated queries, and preserve context throughout interactions. Traditional machine learning (ML) is not able to handle a variety of languages, identify sentiment, or modify responses in real time in response to user behavior, unlike large language models (LLMs). Traditional machine learning models are also less adaptable and scalable because they are usually task-specific and need to be trained separately for new domains. Their efficiency is further reduced by their incapacity to incorporate extensive external knowledge sources or adapt continually from user feedback, leading to less user-friendly, static interactions that don't fulfill the contemporary Quality of experience.}
    \item \textit{Scalability Challenges:} \textcolor{black}{Conventional ML models frequently fail to produce accurate and trustworthy results when there is a lack of training data. These models are prone to overfitting, which occurs when they memorize the small dataset instead of generalizing to new, unseen cases, and they mostly rely on vast quantities of labeled data to learn patterns and generate predictions. Traditional machine ML may perform poorly and have lower predicted accuracy in situations involving sparse data because they are unable to identify the underlying patterns. Furthermore, because it is challenging to extract significant features without a robust dataset, the manual feature engineering approach loses effectiveness when data is insufficient. In these situations, the scalability and adaptability of traditional ML models are significantly restricted by the absence of data.}

\end{itemize}

\begin{figure*}[!t]
	\centering
	\captionsetup{justification=centering}
	\includegraphics[width=14cm, height=14cm]{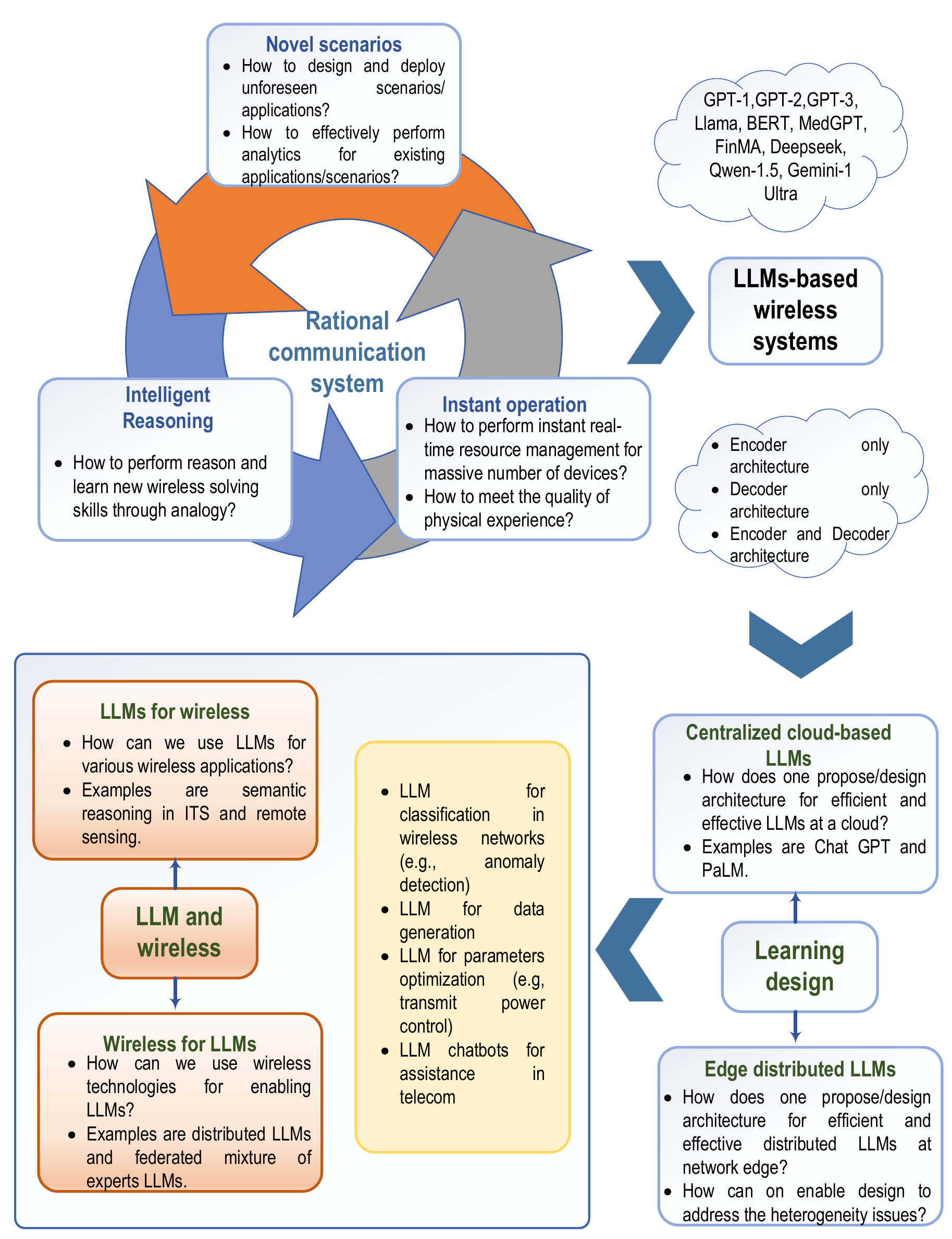}
	\caption{An overview of novel rational wireless, LLMs, and key design aspects.}
	\label{fig:rational_wireless_system}
\end{figure*}

\subsection{A Vision of LLMs-Empowered Rational Wireless Systems}
\subsubsection{LLM-empowered rational wireless networks}
\textcolor{black}{To meet the diverse requirements of end-users in wireless systems, there is a need for a novel system. These requirements will be in terms of quality of service (e.g., latency and reliability) and quality of experience (e.g., user-defined metrics, such as sense of touch in haptics-based healthcare and steering wheel in autonomous cars) \cite {liu2018qoe}. It is very challenging to fulfill the diverse requirements using traditional wireless systems. On the other hand, there will be many novel applications (e.g., brain-computer interaction-based wheelchair for disabled persons) in the foreseeable future. Therefore, it is very challenging to deploy such new and existing applications due to their novel nature and diverse requirements. To implement emerging wireless applications, there is a need for a \emph{rational wireless system} (as shown in Fig.~\ref{fig:rational_wireless_system}) characterized by the following aspects:}      \par

\begin{itemize}
    \item \textit{Novel Scenarios:} \textcolor{black}{In the foreseeable future, a massive number of devices/end-users will be witnessed as per various statistics. Meanwhile, there will many new applications (e.g., fully autonomous cars and haptics-based healthcare) that need effective and efficient deployment. Meanwhile, there will be many applications (e.g., healthcare and smart industry) that have strict latency. For instance, if a patient requests healthcare service from the telehealthcare system, there should be very low latency in serving the patient by the telehealthcare system. Similarly, collision avoidance and accident reporting in autonomous driving cars must be performed with low latency. Moreover, if we consider brain-computer-interaction-enabled smart wheelchairs for disabled persons, they must be served with low latency. On the other hand, reliability must be ensured for various applications. For instance, for healthcare and fully autonomous driving, we must ensure reliability. To fulfill the aforementioned challenges, one must perform analytics before end-users request services from the telecom system. Furthermore, there might be new applications that should be deployed first in digital form. To do so, one must create a virtual model and perform analysis before its deployment. For such deployment, we will need data for intelligent analytics and virtual scenarios. To address these challenges, the works in \cite{han2022dynamic} and \cite{tao2018digital} presented the concept of digital twins and metaverse. However, it is very challenging and difficult to design new scenarios for novel applications that are not yet deployed. Meanwhile, we will need data for intelligent analytics for such scenarios that are not readily available in the traditional metaverse and digital twins-based systems. On the other hand, it is also challenging for the existing wireless systems (e.g., 5G) to meet the design requirements/challenges of the aforementioned applications. }   

    \item \textit{Intelligent Reasoning:} 
     \textcolor{black}{A key component of LLM-native wireless systems is intelligent reasoning, which enables them to comprehend, evaluate, and make sense of learned information. In contrast to traditional AI techniques, intelligent reasoning incorporates cognitive skills including causal reasoning, planning, and analogical reasoning in addition to pattern recognition. This makes it possible for networks to deduce causal linkages, forecast likely outcomes, and adjust to unexpected and changing circumstances. LLMs-native systems can assess complicated, real-world situations, synthesize data into actionable insights, and make decisions on their own that are in line with larger objectives by integrating intelligent thinking. This capacity is essential for tasks like resource allocation optimization, navigating unpredictable situations, and guaranteeing the smooth functioning of digital twins and autonomous agents in a variety of applications, from next-generation wireless networks to metaverse interactions. For instance, the work in \cite{shao2024wirelessllm} enables deductive reasoning (i.e., a type of logical reasoning) by using their LLM models for power allocation using the existing knowledge. }
    \item \textit{Autonomous, Independent Agent-Based Instant Operation:} \textcolor{black}{As explained earlier that there will be a massive number of end-users/devices and other network entities (e.g., edge servers and cloud servers). Additionally, there will be novel performance metrics. For instance, for a haptics-based steering wheel, one can not use traditional metrics (e.g., quality of service in terms of latency). For such applications/scenarios, there should be novel quality of physical experience metrics. Different end-users can have their own physical experience. Therefore, it is very challenging to perform instant resource management for highly dynamic scenarios. Additionally, there will be many management functions (e.g., wireless resource allocation, task offloading, caching decision, transmit power allocation, and association) that need to be performed at end devices. For instance, the work in \cite{nascimento2023self} proposed a MAPE-K model (Monitor, Analyze, Plan, Execute, Knowledge) for self-adaptation in dynamic environments. On the other hand, there will be many applications with a massive number of devices that will be deployed on the same physical infrastructure. This will further make resource management and operation challenging. To resolve this issue, one should propose an autonomous, independent agent-based system design. In such a design, there is a need for a wireless system with many autonomous, independent agents. These agents will perform network management and operation tasks autonomously so as to operate with the least assistance from the network operators.  } 
    \end{itemize}
\subsubsection{Design Aspects}
\textcolor{black}{Regarding the design of LLM-empowered networks, there are two main perspectives: (a) learning algorithm design perspective and (b) wireless perspective \cite{shen2024large}. From a learning algorithm design perspective, one should look at the learning scheme design ranging from selecting the architecture (e.g., transformer-based architecture), number of training parameters (e.g., 180 Billion), attention mechanisms, and number of layers, among others. These parameters should be carefully chosen for a reasonable performance. For LLMs, one can have various architectures, such as transformer-based architecture (e.g., GPT-3 and GPT-4), bidirectional encoder representation from the transformer (RoBERTa), text-to-text transformer, and unified language learning, among others. Mainly, we have two components: (a) encoder and (b) decoder. In the encoder phase, the transformer uses an attention mechanism to reflect the intricate dependencies between various entities (e.g., words in a sentence) for performance improvement compared to traditional convolution neural networks. The operation of the transformer is based on segmenting the input through a tokenization process. These tokens are transformed using the embedded layer into vectors. Transformer uses attention to capture the information between the words in a sequence. Later, normalization and feed-forward neural networks are applied. Therefore, in the encoder phase, the transformer produces a context-rich sequence from an input sequence. In the decoder phase, the input of encoder is used to generate output results. On the other hand, some of the LLMs are only encoder-based (e.g., BERT). Meanwhile, some of the LLMs are only decoder-based (e.g., LLaMA). On the other hand, there is a need for wireless resource optimization to enable efficient learning of LLMs (especially distributed LLMs) over wireless networks. In the next section, we will present a framework based on autonomous independent \emph{Telecom LLM agents}. }

\

\begin{figure*}[!t]
	\centering
	\captionsetup{justification=centering}
	\includegraphics[width=16cm, height=7cm]{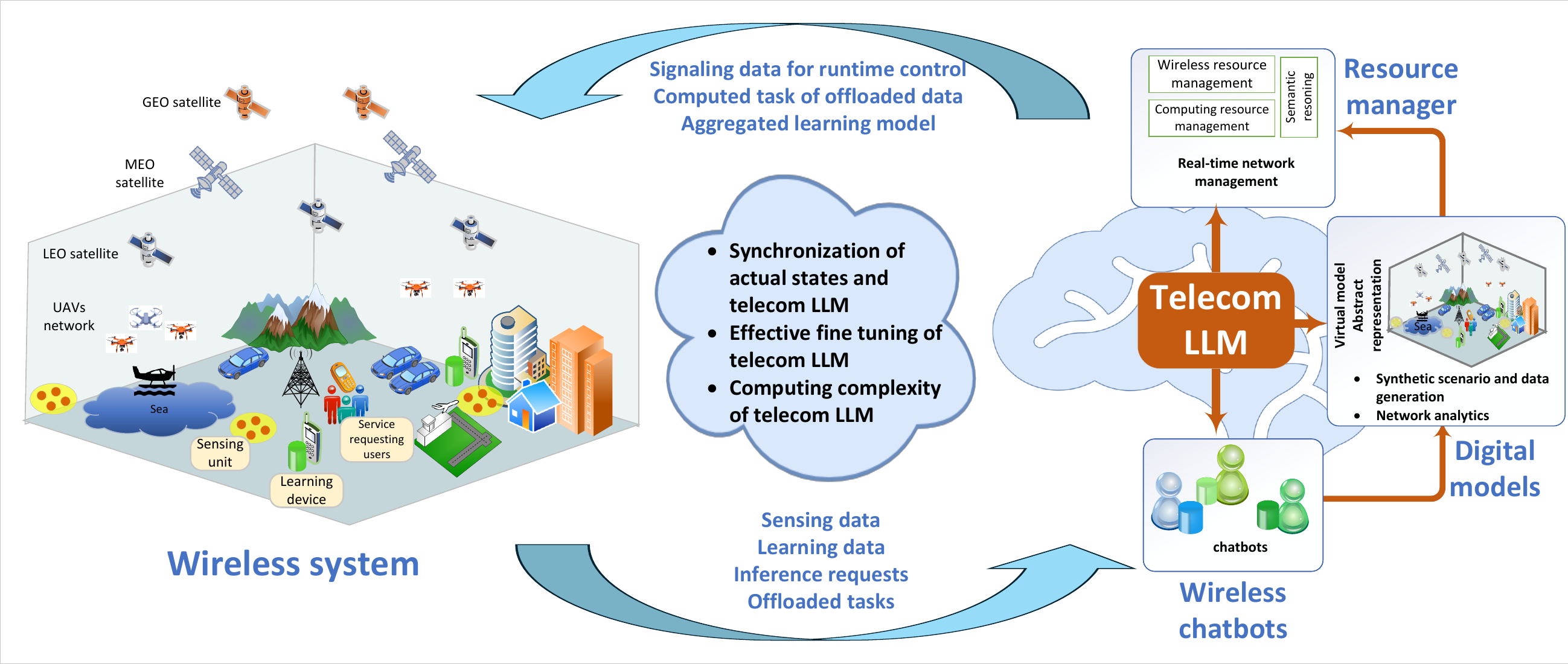}
	\caption{An overview of telecom LLM for IoT.}
	\label{fig:vision}
\end{figure*}

\section{Telecom LLMs: Deployment and Use Case}
\subsection{Framework}
\textcolor{black}{In this section, we present our framework (i.e., as shown in Fig.~\ref{fig:vision}) that is based on telecom LLM agents capable of intelligent reasoning and autonomous independent operation. One can deploy these agents either at the end-devices, edge, or cloud servers. These agents will continuously learn LLM models and perform inferences to serve end-users. An overview of our proposed framework is shown in Fig.~\ref{fig:vision}. Our telecom LLM agent/s will perform mainly three tasks: (a) real-time network management using autonomous operation (e.g., autonomous driving cars), (b) intelligent reasoning using chatbots (e.g., healthcare support and educational tutoring) and inputs from end-users, and (c) analysis of the existing as well as novel scenarios using virtual models and LLMs (e.g., intelligent reasoning for autonomous driving cars). The deployment of telecom LLMs depends mainly on the application requirements. For instance, for inference tasks of remote sensing using vision LLMs, there is a need for more computing power. Therefore, for such kinds of inference tasks, there is a need for the deployment of telecom LLM agents in the remote cloud. \textcolor{black}{In contrast, in applications where computing resource requirements are low and the latency is critical, we need to deploy the telecom LLM agents at the network edge. On the other hand, we can deploy telecom agents both at the network edge and in the cloud to benefit from both cloud and edge.}} \par
\begin{figure*}[!t]
	\centering
	\captionsetup{justification=centering}
	\includegraphics[width=16cm, height=7cm]{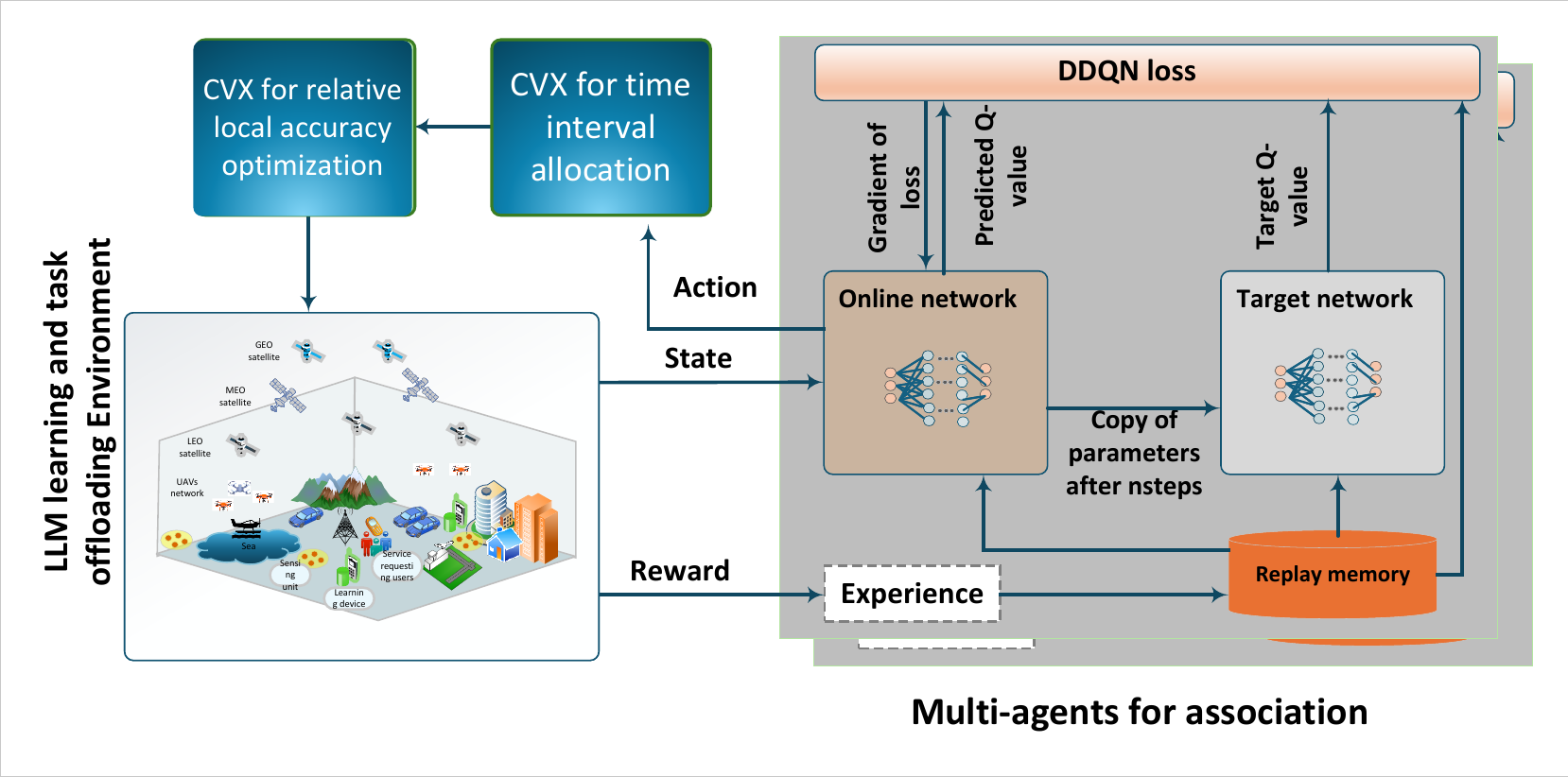}
	\caption{Proposed multi-agent DRL and optimization-based solution.}
	\label{fig:solution}
\end{figure*}

\textcolor{black}{For learning, we adopt federated learning (FL), whereas for inference, the task will be offloaded from end-devices to telecom LLM agents deployed at the network edge. We assume that the LLM model is pre-trained on the remote cloud, and we are fine-tuning the pre-trained model using FL for edge networks. A pre-trained LLM model is fed to all devices at the beginning of learning process. To enable efficient learning of LLMs and inference for end-users, we will need an interaction over wireless networks between end-devices and the edge servers running telecom LLM agents. All the end-devices will learn their local LLMs and then share these local models with the edge running telecom LLMs, where aggregation will take place. Note here that training a local LLM model on devices needs a significant amount of computing. For fine-tuning, we need fewer computing resources to fine-tune the LLM model trained in the cloud. On the other hand, for training a local model for LLM will be challenging and one can take the help of slit FL. Here, we consider fine-tuning of LLM models. \textcolor{black}{Local LLM models are characterized by the relative local accuracy $\theta$, i.e., higher values of $\theta$ reflect low actual local accuracy and vice versa. Note here that $\theta$ has a different interpretation compared to local accuracy. A high value of local accuracy is desirable, whereas a low value of $\theta$ is desirable.} We consider latency as a cost of communication of LLMs learning updates between end-devices and the edge servers. All devices learn local LLM models and then share the trained local models with the edge servers running telecom LLM agents. The telecom LLM agents combine the local LLM models to yield a global LLM model. This global LLM model is sent back to the end-devices for further update. Meanwhile, there will be a set of devices that want their task (e.g., AR rendering task) to be offloaded to the telecom LLMs. Note here that the task offloading requests are less frequent compared to learning LLM update sharing. Meanwhile, the requests should be transmitted to the telecom LLMs with low latency and high reliability. Therefore, we add latency as well as reliability constraints for the task offloading devices. Furthermore, we consider time division multiple access (TDMA) as an access scheme. For TDMA, we assign time slots to various devices in a particular interval. On the other hand, for the association between the end-devices and the telecom LLMs, we follow the unique association of the end-device with a telecom LLM. Meanwhile, the total number of devices associated with a certain telecom LLM should not exceed the maximum serving capacity. Our problem has three optimization variables: (a) association variable (for learning devices and task offloading) $\boldsymbol{a}$, (b) time allocation variable $\tau$, and (c) relative local accuracy $\theta$. This problem is a mixed integer non-linear programming (MINLP) problem. Therefore, we use a solution based on DDQN. We use DDQN for association, whereas for time allocation and relative local accuracy optimization, we use convex optimizers. \textcolor{black}{Fig.~\ref{fig:solution} shows the proposed solution based on DDQN, convex optimization-based time interval allocation, and convex optimization-based relative accuracy optimization. In our problem, the reward depends on task-offloading, time interval allocation, and $\theta$. The task-offloading is performed using DDQN. On the other, we use convex optimizers for solving time interval allocation and relative local accuracy optimization. For DDQN, we are interested in long-term reward maximization, and our problem is actually a cost minimization problem. Therefore, we consider reward as $(\frac{1}{\text{Cost}})$ in our solution.}  }\par
\textcolor{black}{On the other hand, the complexity of our system depends mainly on the DDQN and convex optimizers. Convex optimizers in our system model are used for time interval allocation and $\theta$ optimization. These optimizers have generally low complexity and converge within reasonable iterations \cite{boyd2004convex}. Therefore, we can say that they are suitable for practical implementation. Furthermore, if we consider the complexity of DDQN, it is also reasonable. Mainly, the complexity of DDQN depends on the number of actions and the size of the state space, which are not very high in our system model. Other than the time complexity, deploying agents at end-devices will also need computing resources to train DDQN agents. The computing resources available at end-devices are sufficient to train DDQN agents. However, there is a need for efficient allocation of computing and communication resources for enabling multi-agent reinforcement learning. Overall, we can say that the proposed scheme can be deployed for cost minimization of the distributed LLMs.   }

\begin{figure*}[ht]
    \centering
    \begin{subfigure}[b]{0.32\textwidth} 
        \includegraphics[width=\textwidth]{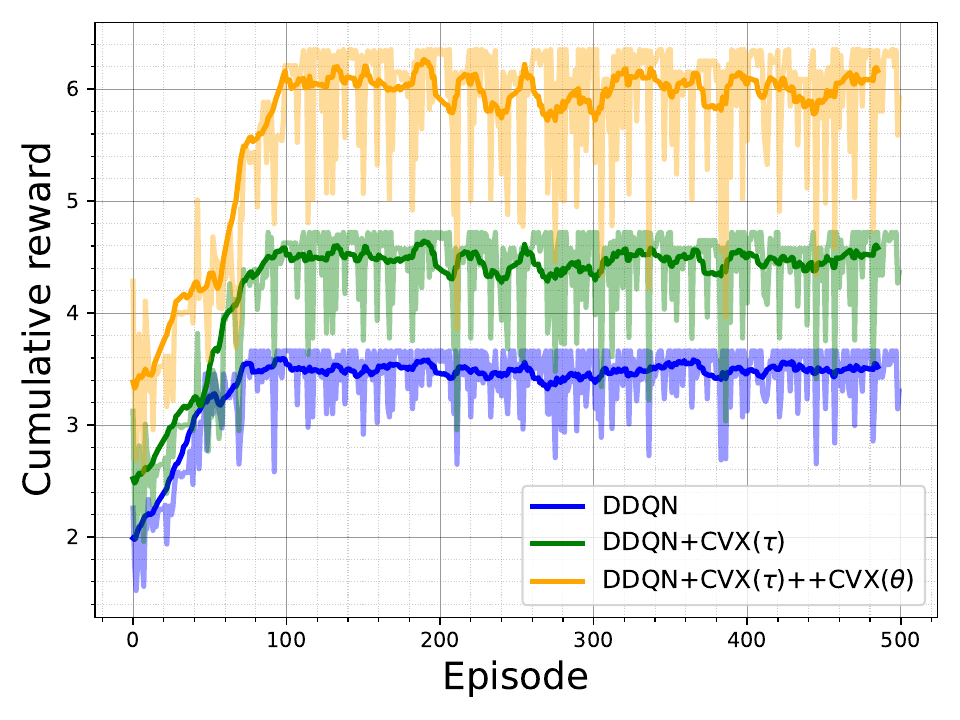}
        \caption{}
        \label{fig:sub1}
    \end{subfigure}
    \hfill
    \begin{subfigure}[b]{0.32\textwidth}
        \includegraphics[width=\textwidth]{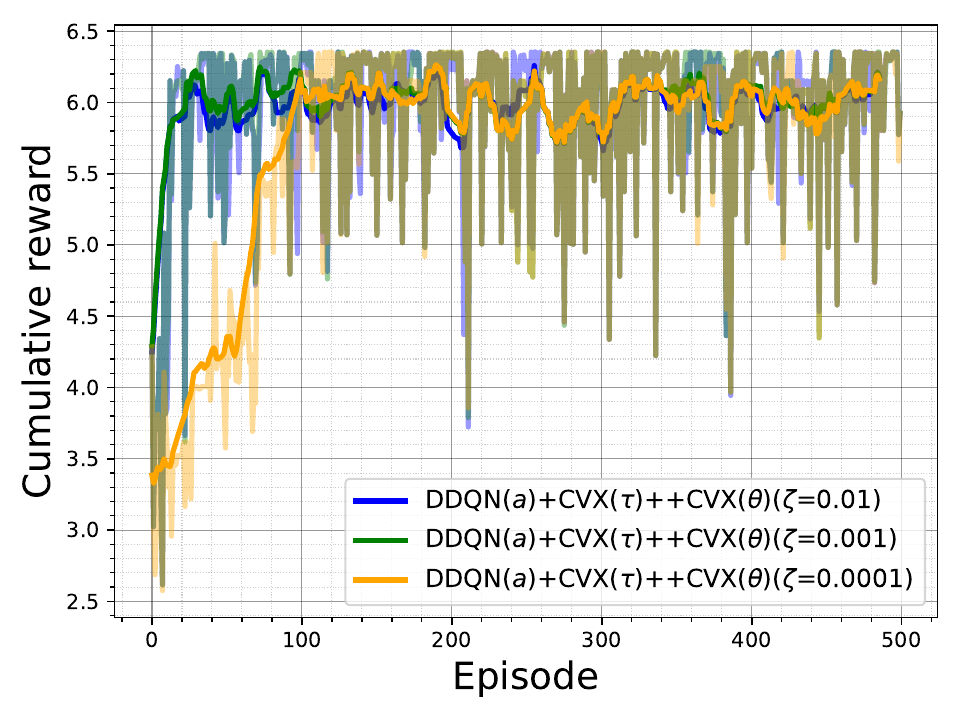}
        \caption{}
        \label{fig:sub2}
    \end{subfigure}
    \hfill
    \begin{subfigure}[b]{0.32\textwidth}
        \includegraphics[width=\textwidth]{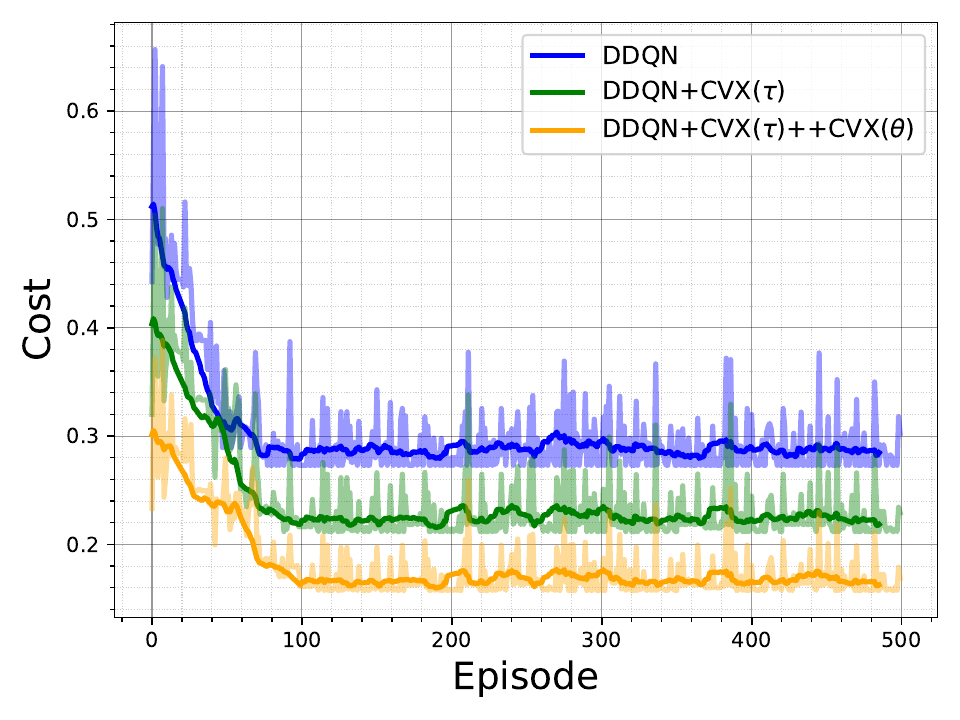}
        \caption{}
        \label{fig:sub3}
    \end{subfigure}
    
    \caption{(a) Reward vs. episodes for various schemes using $36$ agents, $5$ telecom LLM agents deployed at network edge, and $20$ steps, (b) reward vs. episodes for the various learning rates using the proposed scheme for $36$ agents, $5$ telecom LLM agents, and $20$ steps, and (c) cost vs. episodes for various schemes using $36$ agents, $5$ telecom LLM agents deployed at network edge, and $20$ steps. }
    \label{fig:four_figures_33}
\end{figure*}

\subsection{Performance Evaluation}        
\textcolor{black}{We consider an area of $1000 \times 1000m^{2}$ for generation of LLM end-devices. A free space path loss model is considered and devices are assigned power of $50~dBm$. The noise power density is $-174dBm/Hz$. For DDQN, we use a fully connected model that has an input layer, three hidden layers (64,32,32), and an output layer. We use $20$ steps in each episode and a batch size of $16$. An $\epsilon$-greedy strategy is considered with a replay memory size of $500$. Moreover, the discount factor used is $0.9$ and the optimizer used is RMSProp. Fig.~\ref{fig:four_figures_33}a shows the reward vs. episodes for various schemes. We use DDQN in its basic form. Then, we propose BSUM+CVX($\tau$) and DDQN+CVX($\tau$)+CVX($\theta$) as modifications of the DDQN. It is evident from Fig.~\ref{fig:four_figures_33}a that our modified versions of DDQN significantly outperform the traditional DDQN. First, we add a convex optimizer optimizer for time interval allocation due to the convex nature of the time allocation problem. It is clear that after adding a convex optimizer for time allocation, the performance improved significantly. Next, we optimize relative local accuracy, $\theta$, using a convex optimizer. This optimization of $\theta$ further improves the performance in terms of reward. Therefore, we can say that our proposed modified schemes significantly converge fast compared to the traditional DDQN. In Fig.~\ref{fig:four_figures_33}b, we consider DDQN+CVX($\tau$)+CVX($\theta$) and study its performance under various learning rates, $\zeta$. Fig.~\ref{fig:four_figures_33}b shows the stable performance of the proposed scheme for various learning rates. For the learning rate, $\zeta=0.0001$, the convergence is slow compared to other learning rates, $\zeta=0.01$ and $\zeta=0.001$. This is because of slow steps to reach the optimal solution. Finally, we show the cost vs. episodes for various schemes. Again it is clear similar to the previous results (i.e., Figs.~\ref{fig:four_figures_33}a and \ref{fig:four_figures_33}b) that our proposed schemes outperform traditional schemes.            }

\section{Open Challenges}

\subsection{Fine Tuned Multi-Model Edge LLMs}
Implementing multi-model edge LLMs via merging several modalities, including text, pictures, and sensor data is challenging due to the resource limitations of edge devices. Edge environments require effective model optimization strategies to balance latency and performance, in contrast to centralized systems with plenty of processing capacity. Adaptive learning for real-time deployment, energy-efficient inference, and model size reduction via pruning or quantization must all be addressed while maintaining the model's capacity to handle multimodal data efficiently. Furthermore, maintaining data secrecy and synchronizing updates across decentralized nodes without sacrificing accuracy are two more challenges that come with ensuring secure and private fine-tuning in federated settings. To overcome these obstacles and make refined multi-model edge LLMs practical for real-world applications, new methods in distributed training, transfer learning, and resource-efficient architectures are required. For instance, Alpaca is fine-tuned from LLaMA (Large Language Model Meta AI) to improve its conversational abilities and performance in specific tasks. Similarly, one can propose novel, resource optimized, fine-tuned models for wireless networks.

\subsection{Resource Optimized Distributed Edge LLMs}
Using FL to design resource-optimized distributed edge Large Language Models (LLMs) is a multifaceted task. By using locally stored data on edge devices, FL makes decentralized model training possible. This helps protect privacy but also brings with it problems like heterogeneous data distributions, different hardware capabilities, and erratic network connectivity across devices. To effectively optimize LLMs in this situation, sophisticated methods are needed, like adaptive federated optimization algorithms, which minimize communication cost and take device heterogeneity into consideration. FL frameworks must incorporate techniques like model pruning, quantization, and knowledge distillation to minimize resource consumption while preserving the accuracy and generalizability of the LLM. Furthermore, secure aggregation, differential privacy, and strong encryption are necessary to guarantee smooth edge device collaboration while safeguarding sensitive data. 

\subsection{LLMs-Based Self-Sustaining Wireless Networks}
A distinct set of difficulties arises when using Large Language Models (LLMs) for self-sustaining networks because of the intricate relationship between resource limitations, autonomy, and adaptation. LLMs must function in decentralized environments with restricted connection, dynamic topologies, and erratic resource availability in order to support self-sustaining networks, such as those seen in smart cities, disaster recovery areas, or remote IoT devices. Given that these models must process a variety of real-time multimodal data streams while maximizing energy economy and guaranteeing low latency, training and fine-tuning LLMs in such environments is very difficult. Advanced methods in edge-based federated learning, ongoing learning, and model compression are required since the models must also be able to make decisions on their own and learn adaptively without frequently depending on centralized infrastructure. It becomes even more difficult to maintain data security and privacy across dispersed nodes, particularly when managing sensitive data in untrusted settings. Innovative solutions are needed to address these issues and guarantee that LLMs can function consistently and robustly in self-sustaining networks without sacrificing autonomy or performance.

\section{Conclusions}
In this article, we have discussed the limitations of traditional ML schemes in the context of wireless systems. A notion of an LLM-native wireless system is proposed, and a framework is presented for effectively enabling wireless systems through the deployment of distributed telecom LLM agents.. We presented a case study of joint learning and task offloading for telecom LLM agents. For the solution, we proposed a novel scheme based on DDQN along with convex optimization based time interval allocation and relative local accuracy optimization. \textcolor{black}{Our modified solution based on DDQN showed significant performance improvement compared to the traditional DDQN-based solution. We concluded that our work can serve as guidelines for future works on LLMs for wireless networks. We conclude several outcomes from our work. One can combine mathematical optimization schemes with DRL schemes to further improve their performance. Furthermore, LLMs can be deployed in a distributed manner to efficiently enable various wireless applications. Meanwhile, as a future work, one can also optimize the transmit power allocation for further performance enhancement. }

\bibliographystyle{IEEEtran}
\bibliography{Database}

\end{document}